\documentclass[
    10pt,                 
    preprintnumbers,      
    amsmath,              
    amssymb,              
    aps,                  
    longbibliography,     
    unsortedaddress,      
    prf,                  
    floatfix              
]{revtex4-2}

\usepackage{graphicx}       
\usepackage{array}          
\usepackage{booktabs}       
\usepackage{makecell}       
\usepackage{bm}             
\usepackage[normalem]{ulem} 
\usepackage{xcolor}         
\usepackage[colorlinks=true,linkcolor=black,urlcolor=blue,citecolor=red]{hyperref}
\newcommand{\supnum}[1]{\normalfont\textsuperscript{#1}} 
\usepackage{soul}
\usepackage{algorithm,algpseudocode,dsfont}

\begin{document}

\title{Reconstructing Rayleigh--Bénard flows from temperature-only measurements using the Ensemble Kalman Filter}

\author{
Francesco Fossella\supnum{1,2,3~\hyperlink{emailnote}{\textcolor{blue}{*}}},
Luca Biferale\supnum{2,3},
Alberto Carrassi\supnum{4},
Massimo Cencini\supnum{3,5},
Mathieu Desbrun\supnum{6},
Kiwon Um\supnum{1}
}

\affiliation{\vspace{0.15cm}
\mbox{\supnum{1}\textit{LTCI, Telecom Paris, 19 Place Marguerite Perey, 91120 Palaiseau, France.}}\\
\mbox{\supnum{2}\textit{Department of Physics, University of Rome ``Tor Vergata'', Via della Ricerca Scientifica 1, 00133 Rome, Italy.}}\\
\mbox{\supnum{3}\textit{INFN ``Tor Vergata'', Via della Ricerca Scientifica 1, 00133 Rome, Italy.}}\\
\mbox{\supnum{4}\textit{Department of Physics and Astronomy, University of Bologna, Viale Carlo Berti Pichat 6/2, 40127 Bologna, Italy.}}\\
\mbox{\supnum{5}\textit{CNR-ISC, Istituto dei Sistemi Complessi, Via dei Taurini 19, 00185 Rome, Italy.}}\\
\mbox{\supnum{6}\textit{Inria \& École Polytechnique, Route de Saclay, 91120 Palaiseau, France.}}}

\begin{abstract}
Reconstructing the full state of a multiscale chaotic system from observations of only a subset of its degrees of freedom is a central problem in data assimilation. Turbulent Rayleigh--Bénard convection offers a paradigmatic example, for which Charney conjectured that temperature measurements alone should suffice to determine the velocity field. Synchronization-based methods such as nudging and direct replacement nonetheless fail in chaotic regimes, because they require the measurement of the determining modes of the system, a set to which the velocity belongs but the temperature does not. Here we show that the Ensemble Kalman Filter (EnKF), equipped with a proper inflation scheme, removes this asymmetry by exploiting the flow-dependent cross-covariances between temperature and vorticity, and recovers the velocity from noisy and spatio-temporally sparse temperature data alone. We assess the filter in two chaotic regimes of two-dimensional bulk convection by means of a scale-by-scale spectral analysis. When the whole temperature field is observed, both fields synchronize with the true state at all resolved scales, with the temperature estimate becoming even more accurate than the measurements themselves for sufficiently frequent observations. Equispaced observation networks preserve this full-spectrum synchronization down to a threshold density, below which only the largest scales remain synchronized, as also happens when observations are confined to a subdomain. We conclude that Charney’s conjecture is numerically verified in
two-dimensional turbulent convection when the assimilation is
performed using a
properly calibrated EnKF.
\end{abstract}

\maketitle

\section{Introduction}
\label{sec:introduction}
Optimal state estimation from sparse observations remains a central challenge in predictive science~\cite{thompson_uncertainty_1957,talagrand,carrassi_2018,asch_data_assimilation_2016}, especially where chaotic dynamics severely limit predictability horizons~\cite{lorenz_predictability_flow_1969,lorenz96,carrassi_chaotic_dynamics_2022,carrassi_maximum_likelihood_2009}. To address this issue, Pecora and Carroll (1990)~\cite{pecora_1990} introduced the paradigm of \textit{direct replacement} (DR), wherein specific degrees of freedom of a driven system are continuously overwritten by reference ground-truth data. Under a perfect-model scenario, where both systems evolve under identical governing equations but start from different initial conditions, this continuous data insertion alters the Lyapunov spectrum of the driven system, driving the leading conditional Lyapunov exponent negative and forcing complete synchronization~\cite{pikovsky,pazo,carrassi_data_assimilation_2008,pazo_data_assimilation_2016} of the slave system to the master dynamics.

Much of the theoretical foundation for this approach stems from the study of the 2D Navier--Stokes equations~\cite{foias_sur_1967}, which led to the concept of \textit{determining modes}~\cite{olson_titi_determining_modes,carlson_determining_2024,carlson_determining_2025}. These degrees of freedom dictate the asymptotic behavior of the entire flow and, when used in direct replacement or continuous nudging frameworks~\cite{duane_dynamical_2006,nudging}, such as the AOT algorithm (Azouani, Olson, and Titi, 2014)~\cite{azouani_titi}, they enable complete control over the system, achieving full synchronization between all the variables of the assimilating system and those of the ground truth.

Extending these concepts to thermal convection, Charney (1969)~\cite{charney_use_1969} conjectured that observing only the temperature could suffice to reconstruct the full state dynamics, including velocity and pressure, provided that the measurement grid resolves the relevant fine scales. Because thermal convection is a multiscale phenomenon spanning a wide range of scales, capturing fine-scale spatial and spectral information is crucial to avoid decorrelation effects that originate at small scales and prevent synchronization.

Charney's conjecture has been rigorously proven by means of nudging for the 3D planetary geostrophic equations~\cite{farhat_on_2016} and for Rayleigh--Bénard convection in porous media~\cite{farhat_porus_2016}. In standard, fully developed Rayleigh--Bénard turbulence, however, nudging has so far failed to confirm it, as it does not recover the velocity field from temperature observations alone~\cite{altaf2017downscaling,agasthya_rayleigh_2022}. In porous media, Darcy's law eliminates fluid inertia, establishing a direct diagnostic coupling that slaves velocity to temperature variations. In standard turbulent convection, by contrast, buoyancy acts as a continuous energy source across scales, generating fine-scale instabilities that nudging cannot suppress, so that the leading conditional Lyapunov exponent cannot be made negative. This failure reflects a limitation of the method rather than evidence against the conjecture itself.

Synchronization via data assimilation (DA) techniques such as DR or nudging is therefore an asymmetric phenomenon that depends critically on the chosen driving variable. While direct replacement enables complete synchronization of the temperature field when the velocity field is known, attempting to infer the velocity solely from temperature observations fails to control error growth across the turbulent energy cascade, unless the system is in a non-chaotic regime (see \hyperref[app:direct_replacement]{Appendix~\ref*{app:direct_replacement}}).

In this paper, we aim to demonstrate that this asymmetry can be removed by applying a statistical DA framework, such as the Ensemble Kalman Filter (EnKF)~\cite{enkf_original,evensen_book}. We focus on the two-dimensional case, in which the vorticity uniquely determines the velocity field through the stream function: the goal is therefore to infer the vorticity, which remains entirely unobserved, from sparse and noisy measurements of the temperature alone. By exploiting time-dependent cross-correlations between the observed temperature and the unobserved vorticity, we show that the EnKF infers the dynamics across the full range of resolved scales. Through an ensemble-based quantification of the forecast uncertainty, the EnKF propagates information from the observed variables to the unobserved portion of the state vector whenever new data are assimilated, and not solely during the forecast step as in DR and nudging. This allows observational information to propagate across the turbulent energy cascade, circumventing the observability limitations of deterministic methods and enabling the reconstruction of the velocity field from limited temperature data alone.

This paper is organized as follows. \hyperref[sec:rayleigh-benard]{Section~\ref*{sec:rayleigh-benard}} introduces the equations of boundary-free Rayleigh--Bénard convection, summarizes its multiscale nature and scaling properties~\cite{boffetta_two-dimensional_2012,celani2002scaling,winchester_twodimensional_2024,biskamp_scaling_laws}, and presents the simulation setup. \hyperref[sec:enkf]{Section~\ref*{sec:enkf}} details the EnKF framework, including the specific customizations required to model these convective and turbulent flows. \hyperref[sec:enkf_result_overview]{Section~\ref*{sec:enkf_result_overview}} evaluates the performance of the EnKF with either full-field temperature measurements (\hyperref[sub:whole]{Subsec.~\ref*{sub:whole}}) or localized subdomain observations (\hyperref[sub:box]{Subsec.~\ref*{sub:box}}). The analysis focuses on the space-time propagation of information and examines the reconstruction of the temperature together with the multiscale spectral inference of the $\omega$ field. Subsequently, \hyperref[sub:frequency]{Subsec.~\ref*{sub:frequency}} and \hyperref[sub:coarsening]{Subsec.~\ref*{sub:coarsening}} present a systematic investigation of the EnKF performance as a function of the temporal measurement frequency and of the spatial sparseness. Conclusions are drawn in \hyperref[sec:conclusion]{Sec.~\ref*{sec:conclusion}}. \hyperref[app:direct_replacement]{Appendix~\ref*{app:direct_replacement}} presents an energy-based argument showing the failure of direct replacement in the chaotic regime when only the temperature is inserted, and its ability to synchronize when the vorticity is known.

\section{Turbulent Convection without Boundaries}
\label{sec:rayleigh-benard}
The standard Rayleigh--Bénard setup consists of a fluid layer between two horizontal plates, heated from below and cooled from above. The temperature gradient drives convection through a buoyancy-induced nonlinear coupling between the velocity field $\pmb{u}$ and the temperature $T$~\cite{siggia_high_1994,bodenschatz2000recent,lohse2010small,lohse_shishkina_2024}.

Rather than wall-bounded dynamics, we are interested in the statistically homogeneous, fully developed turbulence of the bulk. We therefore consider a doubly periodic square domain and, to sustain convection without physical boundaries, decompose the temperature field as $T(x,z,t) = -\gamma z + \theta(x,z,t)$, with a vertical background gradient $\gamma$ and fluctuations $\theta$~\cite{celani2002scaling,biskamp_scaling_laws,lohse_ultimate_state,winchester_twodimensional_2024}.

Formulated in terms of the out-of-plane vorticity $\omega=\nabla\times\pmb{u}$, the governing equations read
\begin{equation}
\label{eqn:RB}
\begin{cases}
    \partial_t\theta + (\pmb{u} \cdot \nabla) \theta = \gamma u_z + \kappa\nabla^{2} \theta, \\
    \partial_t\omega + (\pmb{u} \cdot \nabla) \omega = g\alpha \frac{\partial \theta}{\partial x} + \nu\nabla^2 \omega- \mu\omega,
\end{cases}
\end{equation}
where $\kappa$ is the thermal diffusivity, $g$ the gravitational acceleration, $\alpha$ the thermal expansion coefficient, and $\nu$ the kinematic viscosity. A friction term $\mu\omega$ is included to dissipate the energy transferred to the largest scales by the inverse kinetic energy cascade~\cite{celani2002scaling}.

The nonlinear exchange between these coupled variables gives rise to multiscale chaotic dynamics characterized by a dual cascade (of temperature variance toward the small scales and of kinetic energy toward the large ones) spanning a wide range of spatial and temporal scales. The balance between inertial and buoyancy forces is set by the Bolgiano--Obukhov (BO) length scale $L_B$~\cite{biskamp_scaling_laws}. Below $L_B$, inertial forces dominate the small-scale dynamics, driving a direct enstrophy cascade, whereas above $L_B$ buoyancy governs an inverse energy cascade towards large scales. In this buoyancy-dominated BO regime, the spectra follow $E_u(k) \sim k^{-11/5}$ (\hyperref[fig:BO]{Fig.~\ref{fig:BO}(a)}) for velocity and $E_\theta(k) \sim k^{-7/5}$ (\hyperref[fig:BO]{Fig.~\ref{fig:BO}(c)}) for temperature~\cite{celani2002scaling,biskamp_scaling_laws}. Beyond these scaling laws, the temperature statistics are nearly Gaussian at large scales and become progressively non-Gaussian and intermittent at small scales (\hyperref[fig:BO]{Fig.~\ref{fig:BO}(d1)}), while the (inverse-cascading) velocity field remains nearly Gaussian and non-intermittent at all scales (\hyperref[fig:BO]{Fig.~\ref{fig:BO}(b1)}).

\subsection{Simulation Setup}
Simulations are performed by integrating \hyperref[eqn:RB]{Eqs.~(\ref{eqn:RB})} on a square domain of side $L=2\pi$ using a pseudospectral scheme with $2/3$-rule dealiasing. The governing equations are integrated in time with an Adams--Bashforth scheme and an exponential integrating factor for the exact integration of the dissipative terms.

We are interested in assessing the performance of data assimilation in reconstructing the flow using only temperature observations with a certain degree of spatial and temporal sparseness. Because such a systematic investigation is numerically expensive within our data assimilation framework (see \hyperref[sec:enkf]{Sec.~\ref{sec:enkf}}), the equations are discretized on a regular grid of $N^2$ points, with $N = 64$ for the systematic parameter sweeps and $N = 256$ for selected experiments. To maximize the Bolgiano--Obukhov scaling range within the resolved scales and ensure a well-defined dissipation cutoff, the standard dissipation operators in \hyperref[eqn:RB]{Eqs.~(\ref{eqn:RB})} are replaced by hyper-viscosity $\nu(-1)^{p+1}\nabla^{2p}\omega$ and hyper-diffusion $\kappa(-1)^{b+1}\nabla^{2b}\theta$, together with a hypo-friction $\mu(-1)^{q+1}\nabla^{-2q}\omega$ that suppresses large-scale energy accumulation~\cite{celani2002scaling}.

Specifically, the $256^2$ setup employs $\nu = \kappa = 3 \times 10^{-15}$, with hyper-viscosity and hyper-diffusivity of order $p = b = 4$, while the $64^2$ setup employs $\nu = \kappa = 5 \times 10^{-21}$, with $p = b = 8$. In both configurations, $\mu = 0.1$, $q = 2$, $\gamma = 1$, and $\alpha g = 0.005$.

\begin{figure*}[t]
    \centering
    \includegraphics[width=\linewidth]{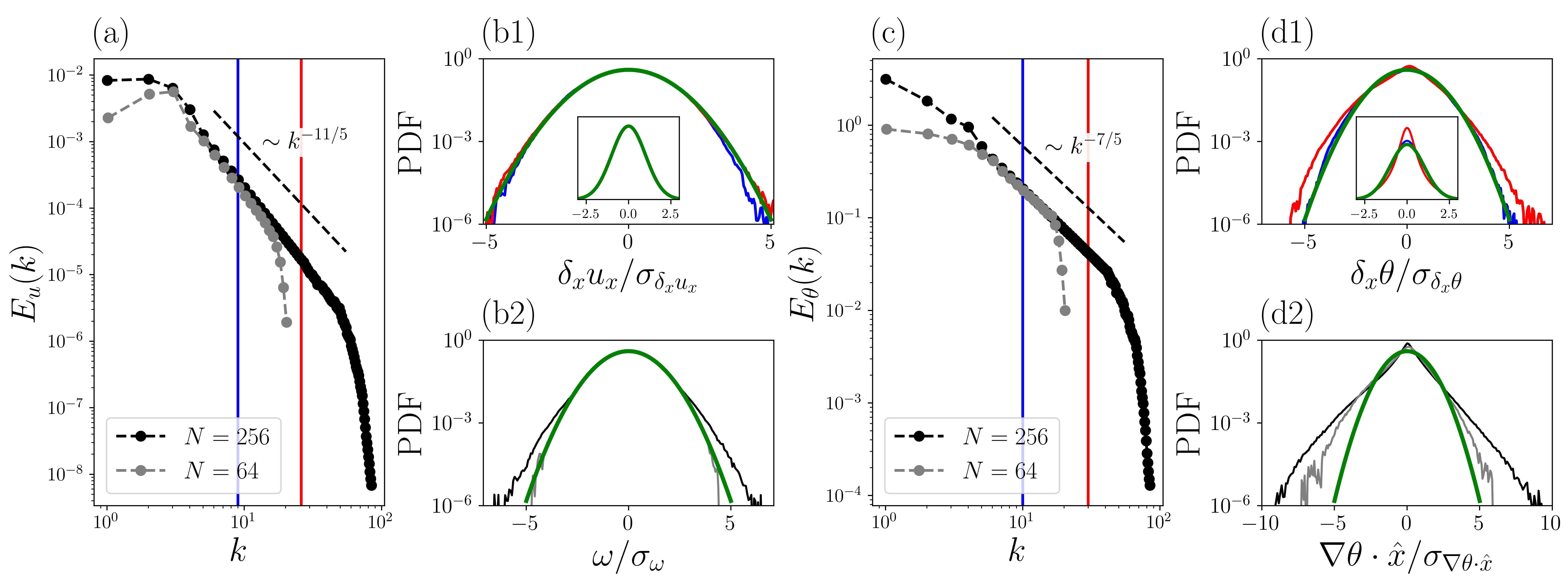}
    \caption{(a) Energy $E_u(k)$ and (c) temperature $E_\theta(k)$ spectra on the $64^2$ and $256^2$ grids, highlighting the Bolgiano--Obukhov range. At shared scales (blue), increment PDFs collapse onto the zero-mean, unit-variance Gaussian (green) for both velocity (b1) and temperature (d1). At the finest scale (red), velocity increments remain nearly Gaussian (b1), whereas temperature develops non-Gaussian features (d1). The linear-scale inset in (d1) shows how thermal intermittency broadens the tails and produces a taller peak, unlike the velocity inset (b1). For the spatial gradients, vorticity $\omega$ (b2) and temperature gradient $\nabla\theta$ (d2), single-point statistics show strongly enhanced non-Gaussian intermittency as finer active scales are resolved on the $256^2$ grid.}
    \label{fig:BO}
\end{figure*}

Both configurations yield an extended Bolgiano--Obukhov range for both velocity (\hyperref[fig:BO]{Fig.~\ref{fig:BO}(a)}) and temperature (\hyperref[fig:BO]{Fig.~\ref{fig:BO}(c)}). As shown in \hyperref[fig:BO]{Fig.~\ref{fig:BO}(b1)}, longitudinal velocity increments exhibit nearly Gaussian, scale-independent statistics, whereas temperature increments display a distinct scale-dependent behavior, with fat tails developing toward smaller scales (\hyperref[fig:BO]{Fig.~\ref{fig:BO}(d1)}). For velocity, the $64^2$ and $256^2$ grids show no qualitative differences in the increment PDFs; for temperature, they agree closely at the large energetic scales, whereas the higher-resolution grid resolves finer spatial structures, thereby capturing enhanced thermal intermittency.

Because differential operators act as high-pass filters sensitive to fine-scale fluctuations, spatial gradients amplify this non-Gaussian behavior. Consequently, the comparison of single-point vorticity PDFs between the $256^2$ and $64^2$ configurations in \hyperref[fig:BO]{Fig.~\ref{fig:BO}(b2)} reveals that higher resolution uncovers sharper localized shear. Similar qualitative trends characterize $\nabla\theta$, with spatial gradients exhibiting progressively stronger non-Gaussian intermittency as more active scales are resolved (\hyperref[fig:BO]{Fig.~\ref{fig:BO}(d2)}).
\section{Ensemble Kalman Filter}
\label{sec:enkf}
The Ensemble Kalman Filter (EnKF) provides a state estimate and an associated error covariance by exploiting cross-correlations between sparse and noisy measured variables and unobserved ones~\cite{enkf_original,evensen_book,carrassi_2018}.

The capabilities of the EnKF in the present setting are investigated using a twin-experiment framework under the perfect-model scenario: the ground truth, from which the synthetic data are sampled, and the model ensemble members evolve under the same governing equations. An ensemble of state realizations, $\hat{\pmb{\phi}}^{(n)}$ ($n = 1, \dots, N_{\text{ens}}$), is integrated forward in time between two successive analyses, i.e., the times at which data are assimilated. In 2D flows, since the out-of-plane vorticity uniquely determines the velocity field, we represent the state as the concatenated vector $\pmb{\phi} = (\pmb{\omega}, \pmb{\theta})^{\rm T}\in\mathbb{R}^{2N^2}$.

When observations become available, the sequential process reaches the \textit{analysis} step~\cite{evensen_book}, at which a correction to each forecast member is computed. Since only the temperature is measured, observations are described by the model
\begin{equation}
\pmb{z}(t_{\text{obs}}) = \pmb{H}\pmb{\theta}(t_{\text{obs}}) + \pmb{\xi}(t_{\text{obs}})\ ,
\end{equation}
where $\pmb{H}\in\mathbb{R}^{M\times N^2}$ is the observation operator, a matrix made of 0s everywhere except for 1s at the $M$ measured grid points of the temperature field, and $\pmb{\xi} \sim \mathcal{N}(\pmb{0}, \pmb{R})$ represents zero-mean observation noise with covariance $\pmb{R}\in\mathbb{R}^{M\times M}$. Among the various EnKF formulations, we adopt the stochastic EnKF, in which each forecast ensemble member $\hat{\pmb{\phi}}^{(n)}$, $n = 1, \dots, N_{\text{ens}}$, is updated into the \textit{analysis} ensemble member $\tilde{\pmb{\phi}}^{(n)}$ according to~\cite{leeuwen}
\begin{equation}
\label{eqn:update}
\tilde{\pmb{\phi}}^{(n)} = \hat{\pmb{\phi}}^{(n)} + \pmb{K} \left[ \pmb{z} - \pmb{H} \hat{\pmb{\theta}}^{(n)} - \pmb{\epsilon}^{(n)} \right]\ ,
\end{equation}
with $\pmb{\epsilon}^{(n)} \sim \mathcal{N}(\pmb{0}, \pmb{R})$ representing synthetic observation perturbations introduced for statistical consistency~\cite{burgers_enkf_analysis}. The Kalman gain $\pmb{K}\in\mathbb{R}^{2N^2\times M}$ is defined as
\begin{equation}
\label{eqn:kalman_gain}
\pmb{K} = \begin{pmatrix} \hat{\pmb{\Sigma}}_{\omega\theta} \\[0.5ex] \hat{\pmb{\Sigma}}_{\theta\theta} \end{pmatrix} \pmb{H}^{\rm T} \left( \pmb{H} \hat{\pmb{\Sigma}}_{\theta\theta} \pmb{H}^{\rm T} + \pmb{R} \right)^{-1},
\end{equation}
where the forecast cross-covariance between vorticity and temperature, $\hat{\pmb{\Sigma}}_{\omega\theta}$, and the forecast temperature covariance, $\hat{\pmb{\Sigma}}_{\theta\theta}$,\footnote{By omitting the explicit velocity components from $\pmb{\phi}$, we keep the dimension of the covariance matrices minimal, avoiding prohibitively large matrices while fully preserving the thermal-fluid coupling.} are estimated from the ensemble as
\begin{equation}
\label{eqn:covariances}
\hat{\pmb{\Sigma}}_{\omega\theta}=\frac{1}{N_{\text{ens}}-1}\sum_{n=1}^{N_{\text{ens}}} \big(\hat{\pmb{\omega}}^{(n)} - \langle\hat{\pmb{\omega}}\rangle_{N_{\text{ens}}}\big)\big(\hat{\pmb{\theta}}^{(n)} - \langle\hat{\pmb{\theta}}\rangle_{N_{\text{ens}}}\big)^{\rm T}, \qquad
\hat{\pmb{\Sigma}}_{\theta\theta}=\frac{1}{N_{\text{ens}}-1}\sum_{n=1}^{N_{\text{ens}}} \big(\hat{\pmb{\theta}}^{(n)} - \langle\hat{\pmb{\theta}}\rangle_{N_{\text{ens}}}\big)\big(\hat{\pmb{\theta}}^{(n)} - \langle\hat{\pmb{\theta}}\rangle_{N_{\text{ens}}}\big)^{\rm T},
\end{equation}
with $\langle\cdot\rangle_{N_{\text{ens}}} = \frac{1}{N_{\text{ens}}}\sum_{n=1}^{N_{\text{ens}}}(\cdot)$ denoting the ensemble average. The vorticity covariance $\hat{\pmb{\Sigma}}_{\omega\omega}$, used below for the inflation, is defined analogously.

The Kalman gain (\hyperref[eqn:kalman_gain]{Eq.~(\ref{eqn:kalman_gain})}) balances observations against forecast predictions, weighting them according to the relative magnitude of the measurement uncertainty $\pmb{R}$ and of the forecast uncertainty of the temperature at the measured points, $\pmb{H}\hat{\pmb{\Sigma}}_{\theta\theta}\pmb{H}^{\rm T}$. The block $\hat{\pmb{\Sigma}}_{\theta\theta}$ then corrects the observed temperature, while the cross-covariance block $\hat{\pmb{\Sigma}}_{\omega\theta}$ transfers the correction to the unobserved vorticity.

At the end of each assimilation cycle, the newly computed analysis fields serve as initial conditions to advance the ensemble through the next forecast phase. When proper observability conditions are met~\cite{bocquet_ensemble_variational_2017}, the sequential alternation of forecast and analysis drives the ensemble spread to a quasi-stationary regime, where the assimilation of new observations balances intrinsic model instabilities and error growth~\cite{carrassi_data_assimilation_2008}. In this regime, the state uncertainty reaches a statistical floor, and the filter tracks the ground truth as closely as the available information allows.

In our 2D convection application, both the ground truth and the ensemble are evolved using a pseudospectral solver, enforcing a zero spatial mean on all fields and applying the $2/3$ rule to prevent aliasing errors. However, because the stochastic analysis update
\hyperref[eqn:update]{(\ref{eqn:update})} can reintroduce energy at wavenumbers beyond the dealiasing cutoff, dealiasing is explicitly reapplied after every analysis step, before the subsequent forecast phase.

\subsection{Two-Field Inflation}
The stability and efficiency of Ensemble Kalman Filter algorithms are constrained by sampling errors arising from finite ensemble sizes~\cite{evensen_book}. When the ensemble is too small to adequately resolve the spectrum of unstable and neutral Lyapunov directions~\cite{grudzien_covariance_inflation,carrassi_chaotic_dynamics_2022}, the resulting covariance estimates become prone to an underestimation of the variances (diagonal elements) and to spurious long-distance correlations (off-diagonal elements). These often cause nonphysical and dynamically unbalanced state corrections that contaminate the dynamics. Spurious long-distance correlations are usually, and often effectively, counteracted by {\it localization}~\cite{carrassi_2018,evensen_book}. Although localization is almost unavoidable in high-dimensional systems, we do not apply it here, since well-established schemes damp long-range correlations within each field~\cite{gaspari_cohn_construction_1999} but offer no clear prescription for the cross-correlations between the two fields, through which the assimilation propagates from the measured to the unmeasured field.

We found it unavoidable, however, to use {\it inflation}, the other well-known remedy adopted in EnKFs to counteract the tendency of the ensemble to collapse and thus to mitigate or, in the best case, prevent filter divergence~\cite{localization2,enkf_practical,grudzien2018}. As the ensemble collapses, the filter over-relies on the forecast: the Kalman gain (\hyperref[eqn:kalman_gain]{Eq.~(\ref{eqn:kalman_gain})}) decays, and subsequent observations effectively lose their influence. To restore this information flow, we implement a dynamic inflation~\cite{inflation,raanes_adaptive_inflation_2019} technique that rescales each analysis ensemble member $(\tilde{\pmb{\theta}}^{(n)}, \tilde{\pmb{\omega}}^{(n)})$ around the ensemble mean:
\begin{equation}
\label{eqn:inflation}
\begin{aligned}
\tilde{\pmb{\theta}}^{(n)} &\leftarrow \langle\tilde{\pmb{\theta}}\rangle_{N_{\text{ens}}} + g_{\theta}(t)\left(\tilde{\pmb{\theta}}^{(n)} - \langle\tilde{\pmb{\theta}}\rangle_{N_{\text{ens}}}\right), \\[1ex]
\tilde{\pmb{\omega}}^{(n)} &\leftarrow \langle\tilde{\pmb{\omega}}\rangle_{N_{\text{ens}}} + g_{\omega}(t)\left(\tilde{\pmb{\omega}}^{(n)} - \langle\tilde{\pmb{\omega}}\rangle_{N_{\text{ens}}}\right),
\end{aligned}
\end{equation}
where $g_\theta(t)$ and $g_\omega(t)$ are scalar coefficients applied uniformly to all grid points of the corresponding field. This operation preserves the ensemble mean and rescales the analysis covariances as $\tilde{\pmb{\Sigma}}_{\theta\theta}\to g_\theta^2\,\tilde{\pmb{\Sigma}}_{\theta\theta}$ and $\tilde{\pmb{\Sigma}}_{\omega\omega}\to g_\omega^2\,\tilde{\pmb{\Sigma}}_{\omega\omega}$. The dynamic inflation coefficients are defined as
\begin{equation}
\begin{aligned}
\label{eqn:inflation_coeff}
g_\theta(t) &= 1 + \eta \left| \frac{\text{tr}(\hat{\pmb{\Sigma}}_{\theta\theta}) - \text{tr}(\tilde{\pmb{\Sigma}}_{\theta\theta})}{\text{tr}(\hat{\pmb{\Sigma}}_{\theta\theta})} \right| ,\\[1ex]
g_\omega(t) &= 1 + \eta \left| \frac{\text{tr}(\hat{\pmb{\Sigma}}_{\omega\omega}) - \text{tr}(\tilde{\pmb{\Sigma}}_{\omega\omega})}{\text{tr}(\hat{\pmb{\Sigma}}_{\omega\omega})} \right|.
\end{aligned}
\end{equation}
By computing the traces of the forecast ($\hat{\pmb{\Sigma}}$) and analysis ($\tilde{\pmb{\Sigma}}$) covariance matrices separately for each field, we apply a variable-dependent inflation that respects the distinct physical scales of vorticity and temperature. Specifically, we use a common prefactor $\eta$ combined with the respective field ratios to generate the different required inflations, with $\eta$ acting as the regulator that balances model fidelity against the observational update. This mechanism counteracts the collapse of the ensemble spread, measured by $\text{tr}(\tilde{\pmb{\Sigma}})$, by promoting a controlled relaxation toward the forecast spread, $\text{tr}(\hat{\pmb{\Sigma}})$. Our approach is inspired by the relaxation-to-prior-spread (RTPS) inflation method used in several implementations of the EnKF for numerical weather prediction~\cite{RTPS,kotsuki,tandeo_2020}. By making the inflation proportional to the magnitude of the spread reduction, we prevent the system from becoming overconfident in its forecast, thereby preserving a nonzero Kalman gain and ensuring that observational data remain integral to the assimilation process throughout the entire integration. In all experiments, the prefactor $\eta$ was tuned through extensive testing.

\section{Results}
\label{sec:enkf_result_overview}
The experimental framework evaluates the EnKF by evolving the ground truth and the ensemble realizations under identical forward dynamics from distinct initial conditions (twin experiment within the perfect-model scenario). 

To assess how spatial information is reconstructed and propagated across the domain, we first focus on the high-resolution $256^2$ configuration (see \hyperref[sec:rayleigh-benard]{Sec.~\ref{sec:rayleigh-benard}}). We compare two distinct observational setups: assimilating (i) the temperature everywhere in the domain (\hyperref[sub:whole]{Subsec.~\ref{sub:whole}}), which provides the reference benchmark, and (ii) the temperature within a subdomain (\hyperref[sub:box]{Subsec.~\ref{sub:box}}), so as to trace how localized observations can inform the full-domain reconstruction. Theoretically, the ensemble size $N_{\text{ens}}$ must scale with the dimension of the unstable and neutral Lyapunov subspace~\cite{carrassi_2018, bocquet_degenerate_2017,chen_inferring_2021}. Although a rigorous computation of the full Lyapunov spectrum lies beyond the scope of this work, numerical evidence indicates that an ensemble of $N_{\text{ens}} = 10^4$ members with inflation is sufficient to resolve the active instabilities and ensure the convergence of the filter.

The appropriate time interval between consecutive analyses is physically tied to the system dynamics. Because decorrelation between the ground truth and the ensemble members is driven by the fast, small scales, the assimilation frequency must be related to the characteristic timescale of these scales, which we estimate through the Lyapunov time $t_{\lambda}=1/\lambda$, where $\lambda$ is the leading Lyapunov exponent computed as in~\cite{boffetta_musacchio_chaos_predictability_2017}; in the $256^2$ case, $t_{\lambda}=4.5$. Systematic analyses performed at lower resolution ($64^2$; see below) suggest that reasonably good tracking is achieved when the observation interval is $\Delta t_{\text{obs}}=t_{\lambda}/8 \approx 0.5$.

The measurement error covariance matrix is chosen to be diagonal and stationary in time, $R(\pmb{r}, \pmb{r}') = \sigma^2 \delta(\pmb{r} - \pmb{r}')$, corresponding to uniform variance and no spatial correlation. The noise level $\sigma$ is calibrated to perturb only the high-wavenumber end of the spectrum.

While the $256^2$ setup provides insight into the spatial transfer of information in a strongly turbulent regime, a broad multi-parameter sweep at this resolution would be computationally prohibitive. Such a sweep involves varying the observation frequency when measuring the entire temperature field (\hyperref[sub:frequency]{Subsec.~\ref{sub:frequency}}) and changing the density of measurement points while holding the observation interval fixed at $\Delta t_{\text{obs}} = t_{\lambda}/8$ (\hyperref[sub:coarsening]{Subsec.~\ref{sub:coarsening}}). For this reason, the systematic parameter sweeps are conducted on the reduced $64^2$ system. Owing to the smaller number of unstable Lyapunov directions in this regime, an ensemble of $N_{\text{ens}}=2000$ members with inflation is sufficient to avoid sampling degradation. For this system, the Lyapunov time is $t_{\lambda} = 7.7$, and the observation intervals are expressed as fractions of this value.

Note that the $256^2$ and $64^2$ configurations differ not merely in spatial resolution: they represent two distinct chaotic regimes, with different numbers of active degrees of freedom and different attractor dimensions. Testing the EnKF in both setups allows us to evaluate its robustness under different dynamical complexities.

For all experimental configurations, prior to data assimilation, the ensemble undergoes a spin-up phase in which it evolves freely from random initial states until it converges onto the chaotic attractor. This step guarantees that the initial ensemble members correctly sample the invariant measure of the system and develop physically realistic cross-covariances, with an initial spread bounded by the size of the attractor~\cite{enkf_practical,carrassi_maximum_likelihood_2009}. Finally, all assimilation experiments are integrated over a total duration $T_{\text{tot}}=250$, corresponding to about $8\,t_L$, where $t_L$ is the large-scale turnover time, so that all performance metrics reflect a statistically stationary regime.

The complete pipeline used to perform the experiments is outlined in \hyperref[alg:enkf]{Algorithm~\ref{alg:enkf}}.

\begin{algorithm}[H]
\caption{Stochastic EnKF with inflation}
\label{alg:enkf}
\begin{algorithmic}[1]
\State $\pmb{\phi}(t_0)=(\pmb{\omega}(t_0),\pmb{\theta}(t_0))$
    \Comment{Thermalize ground truth}
\State $\hat{\pmb{\phi}}^{(n)}(t_0)=(\hat{\pmb{\omega}}^{(n)}(t_0),\hat{\pmb{\theta}}^{(n)}(t_0))$
    \Comment{Thermalize ensemble}
\For{$j=1,\dots,T_{\text{tot}}/dt$}
    \Comment{$t_j=t_0+j\,dt$ ($dt$ integration step)}
    \State $\pmb{\phi}(t_j)=(\pmb{\omega}(t_j),\pmb{\theta}(t_j))$
        \Comment{Evolve ground truth}
    \State $\hat{\pmb{\phi}}^{(n)}(t_j)=(\hat{\pmb{\omega}}^{(n)}(t_j),\hat{\pmb{\theta}}^{(n)}(t_j))$
        \Comment{Forecast ensemble}
    \If{$\mathrm{mod}(j,n_{\text{obs}})=0$}
        \Comment{$n_{\text{obs}}=\Delta t_{\text{obs}}/dt$}
        \State $\hat{\pmb{\Sigma}}_{\omega\theta}=\frac{1}{N_{\text{ens}}-1}\sum_{n=1}^{N_{\text{ens}}}\big(\hat{\pmb{\omega}}^{(n)}-\langle\hat{\pmb{\omega}}\rangle_{N_{\text{ens}}}\big)\big(\hat{\pmb{\theta}}^{(n)}-\langle\hat{\pmb{\theta}}\rangle_{N_{\text{ens}}}\big)^{\rm T}$
        \Comment{Cross-covariance\hyperref[eqn:covariances]{~(\ref{eqn:covariances})}}
        \State $\hat{\pmb{\Sigma}}_{\theta\theta}=\frac{1}{N_{\text{ens}}-1}\sum_{n=1}^{N_{\text{ens}}}\big(\hat{\pmb{\theta}}^{(n)}-\langle\hat{\pmb{\theta}}\rangle_{N_{\text{ens}}}\big)\big(\hat{\pmb{\theta}}^{(n)}-\langle\hat{\pmb{\theta}}\rangle_{N_{\text{ens}}}\big)^{\rm T}$
        \Comment{Temperature covariance\hyperref[eqn:covariances]{~(\ref{eqn:covariances})}}
        \State $\pmb{K}=\begin{pmatrix}\hat{\pmb{\Sigma}}_{\omega\theta}\\ \hat{\pmb{\Sigma}}_{\theta\theta}\end{pmatrix}\pmb{H}^{\rm T}(\pmb{H}\hat{\pmb{\Sigma}}_{\theta\theta}\pmb{H}^{\rm T}+\pmb{R})^{-1}$
            \Comment{Kalman gain\hyperref[eqn:kalman_gain]{~(\ref{eqn:kalman_gain})}}
        \State $\tilde{\pmb{\phi}}^{(n)}=\hat{\pmb{\phi}}^{(n)}+\pmb{K}\,(\pmb{z}-\pmb{H}\hat{\pmb{\theta}}^{(n)}-\pmb{\epsilon}^{(n)})$
            \Comment{Analysis\hyperref[eqn:update]{~(\ref{eqn:update})}}
        \State $\langle\tilde{\pmb{\omega}}\rangle_{N_{\text{ens}}}+g_\omega\,(\tilde{\pmb{\omega}}^{(n)}-\langle\tilde{\pmb{\omega}}\rangle_{N_{\text{ens}}})\to\tilde{\pmb{\omega}}^{(n)}$
            \Comment{Inflate\hyperref[eqn:inflation]{~(\ref{eqn:inflation})}}
        \State $\langle\tilde{\pmb{\theta}}\rangle_{N_{\text{ens}}}+g_\theta\,(\tilde{\pmb{\theta}}^{(n)}-\langle\tilde{\pmb{\theta}}\rangle_{N_{\text{ens}}})\to\tilde{\pmb{\theta}}^{(n)}$
            \Comment{Inflate\hyperref[eqn:inflation]{~(\ref{eqn:inflation})}}
        \State $\mathit{Dealias}\,(\tilde{\pmb{\omega}}^{(n)},\tilde{\pmb{\theta}}^{(n)})\to(\tilde{\pmb{\omega}}^{(n)},\tilde{\pmb{\theta}}^{(n)})$
            \Comment{$2/3$ rule}
        \State $\hat{\pmb{\phi}}^{(n)}(t_j)=\tilde{\pmb{\phi}}^{(n)}(t_j)$
            \Comment{Reset prior on analysis}
    \EndIf
\EndFor
\end{algorithmic}
\end{algorithm}

\begin{figure*}[t]
    \centering
    \includegraphics[width=\linewidth]{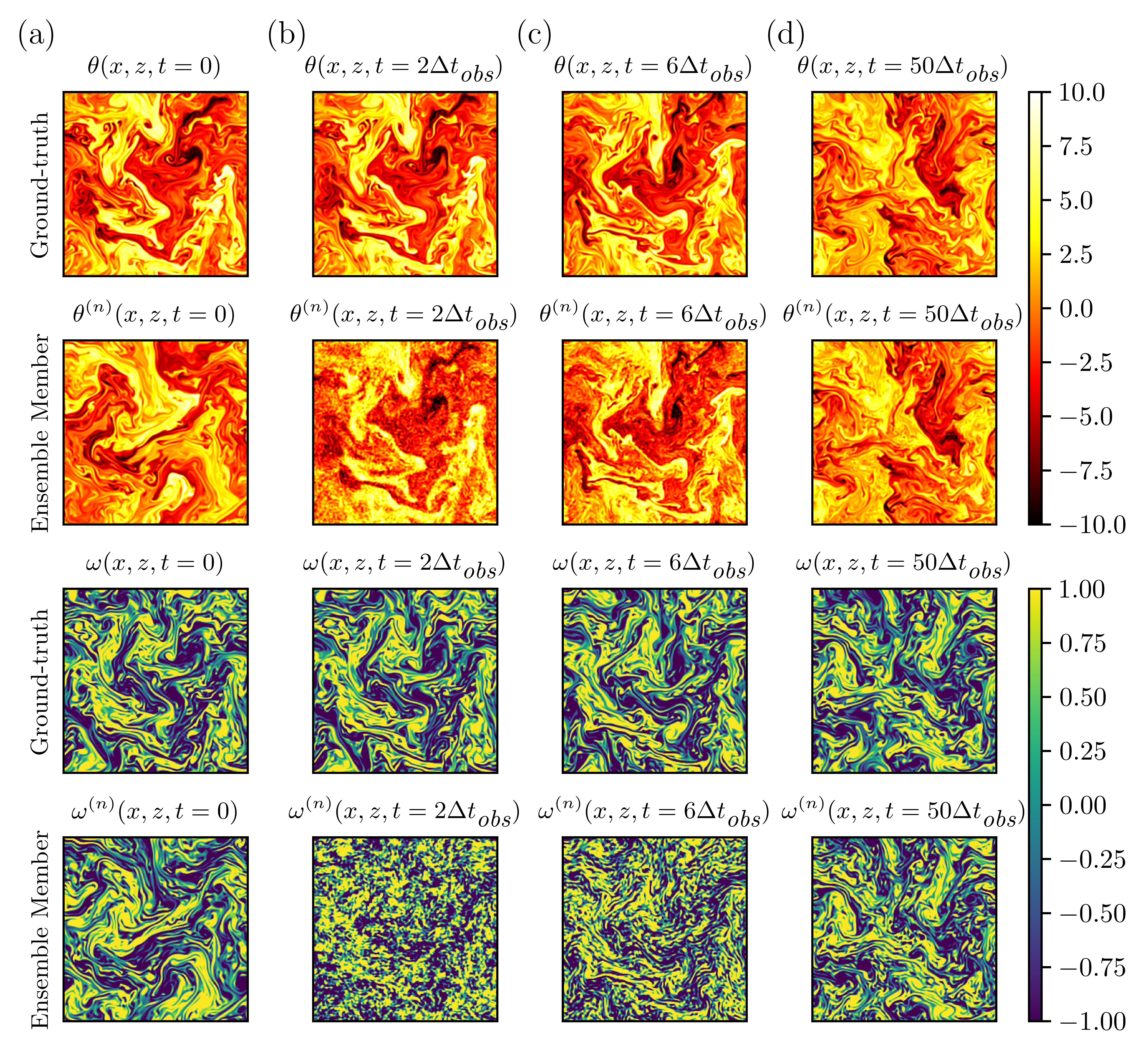}
    \caption{Time evolution of the ground-truth temperature and vorticity (first and third rows) compared with a randomly selected ensemble member (second and fourth rows), both initialized on the attractor, on a $256^2$ grid with measurements over the whole domain: (a) initial independent states; (b) states after two measurements, showing partial large-scale synchronization of the temperature, with persistent small-scale noise, and an ongoing correction of the vorticity; (c) states after six measurements, showing a well-assimilated temperature, while the vorticity begins to match the ground truth at the large scales; (d) states after 50 measurements, showing a qualitatively complete assimilation of the temperature and a nearly complete assimilation of the vorticity, with minimal residual small-scale noise.}
    \label{fig:measuring_whole_fields}
\end{figure*}
\medskip
\noindent\textbf{Error Metrics}\par
\noindent
To evaluate how well the filter assimilates the ground-truth state across scales, we introduce metrics based on the energy spectrum. The energy spectrum of a generic spatial field $\phi(x, z, t)$ is defined as
\begin{equation}
    E_\phi(k,t) = \sum_{\pmb{k} \in S_k} |\phi(\pmb{k}, t)|^2,
\end{equation}
where $\phi(\pmb{k}, t)$ denotes the Fourier mode at wavevector $\pmb{k}$, and $S_k$ the shell centered at wavenumber $k$. We then define the error spectra for the temperature and vorticity fields of the $n$-th ensemble member as $E_{\delta\theta}^{(n)}(k,t)=\sum_{\pmb{k} \in S_k} |\theta(\pmb{k},t)-\theta^{(n)}(\pmb{k},t)|^2$ and $E_{\delta\omega}^{(n)}(k,t)=\sum_{\pmb{k} \in S_k} |\omega(\pmb{k},t)-\omega^{(n)}(\pmb{k},t)|^2$, respectively. Note that no explicit superscript is used here to distinguish the forecast ($\hat{\cdot}$) from the analysis ($\tilde{\cdot}$) state, as this formulation describes the general trajectory of a member during its temporal evolution and subsequent correction. To analyze the long-term statistical behavior of the filter, these spectra are averaged in time within the saturation regime of the EnKF: $E_{\delta\theta}^{(n)}(k)=\langle E_{\delta\theta}^{(n)}(k,t)\rangle_T=\frac{1}{T}\int_{t_s}^{t_s+T} E_{\delta\theta}^{(n)}(k,t)\,\mathrm{d}t$, where $t_s$ marks the onset of the stationary regime and $T$ is the length of the averaging window.
\begin{figure*}[t]
    \centering
    \includegraphics[width=\linewidth]{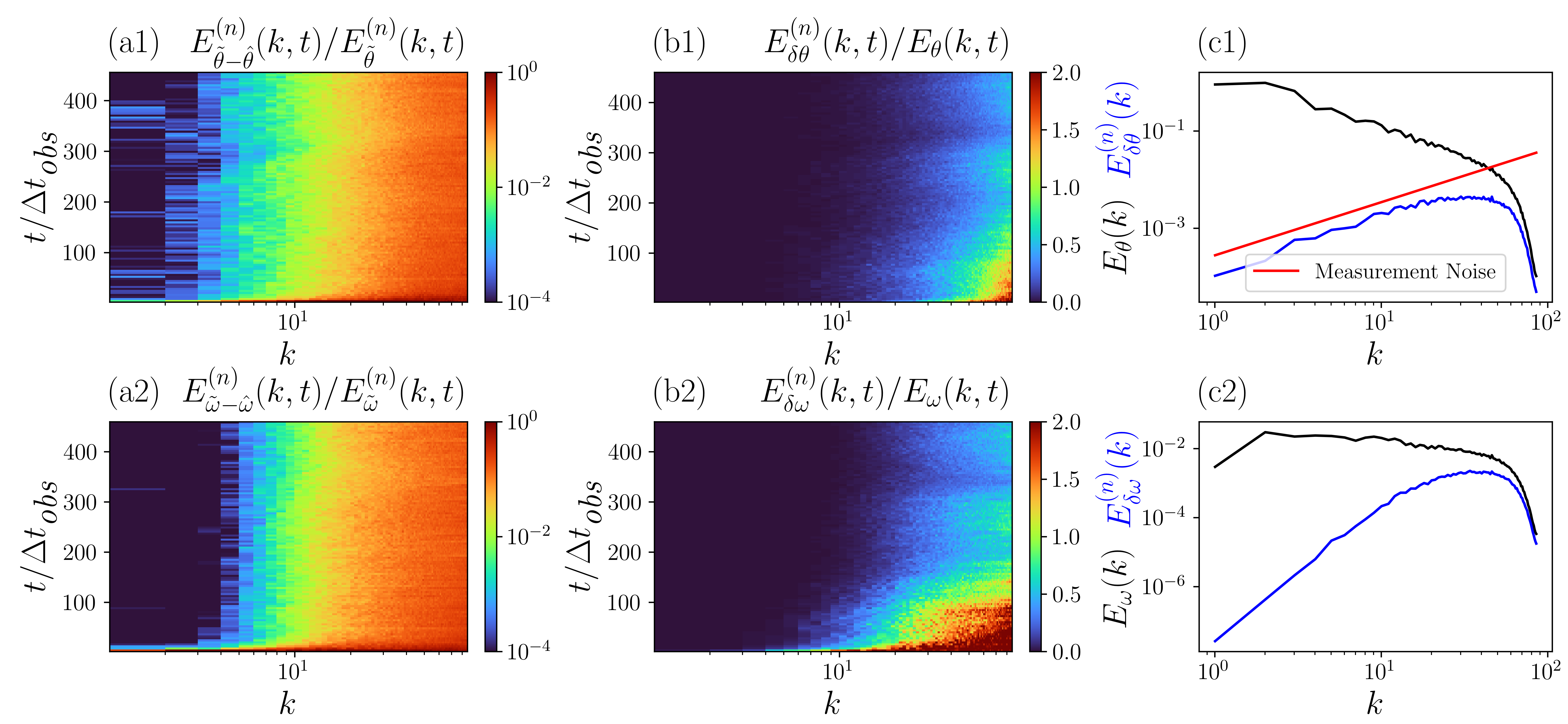}
    \caption{Normalized spectral corrections applied by the EnKF to the $n$-th member for temperature (a1) and vorticity (a2), showing the scales modified during the data assimilation routine when the entire temperature field is measured on a $256^2$ grid. Panels (b1) and (b2) display the corresponding normalized spectral errors for both fields. Panels (c1) and (c2) present time-averaged results over the stationary saturation regime, showing complete synchronization at all resolved scales for both fields. The red curve shows the spectrum of the measurement error, which, as discussed in the main text, mainly affects the small scales.}
    \label{fig:corrections_whole_fields}
\end{figure*}
Another key quantity is the relative spectral correction induced by the assimilation process on the $n$-th ensemble member, defined as
\begin{equation}
\label{eqn:rel_corection}
    \frac{E^{(n)}_{\tilde{\phi} - \hat{\phi}}(k,t)}{E^{(n)}_{\tilde{\phi}}(k,t)} = \frac{\sum_{\pmb{k} \in S_k} \left| \tilde{\phi}^{(n)}(\pmb{k},t) - \hat{\phi}^{(n)}(\pmb{k},t) \right|^2}{\sum_{\pmb{k} \in S_k} \left| \tilde{\phi}^{(n)}(\pmb{k},t) \right|^2}\ ,
\end{equation}
where $\tilde{\pmb{\phi}}^{(n)}-\hat{\pmb{\phi}}^{(n)}=\pmb{K}\left[\pmb{z}-\pmb{H}\hat{\pmb{\theta}}^{(n)}-\pmb{\epsilon}^{(n)}\right]$ is the update increment produced by the EnKF (\hyperref[eqn:update]{Eq.~(\ref{eqn:update})}).

\subsection{Measuring the Whole Domain}
\label{sub:whole}
The primary objective of this experiment is to reconstruct the temperature field and recover the unobserved vorticity through a multiscale information cascade by observing only the full temperature field. For this case, an observation interval $\Delta t_{\text{obs}} = t_{\lambda}/8$ is selected and, following extensive testing, an inflation prefactor $\eta = 0.55$ is applied (see \hyperref[eqn:inflation_coeff]{Eq.~(\ref{eqn:inflation_coeff})}). The noise level of the diagonal measurement error covariance is set to $\sigma = 3$, so as to perturb only the high-wavenumber end of the spectrum (see the red curve in \hyperref[fig:corrections_whole_fields]{Fig.~\ref{fig:corrections_whole_fields}(c1)}).

Although only temperature observations are available and the dynamics is fully developed and chaotic, the EnKF corrects large initial-condition errors in both fields. \hyperref[fig:measuring_whole_fields]{Figure~\ref{fig:measuring_whole_fields}} shows this transition through the time evolution of the ground-truth temperature and vorticity alongside a randomly chosen ensemble member. Initially (\hyperref[fig:measuring_whole_fields]{column (a)}), the ground-truth and member fields are independent and uncorrelated, although both reside on the attractor. After two measurements (\hyperref[fig:measuring_whole_fields]{column (b)}), the member temperature is synchronized with the ground truth at the large scales, while retaining small-scale noise, whereas the vorticity remains uncertain and undergoes large corrections at all scales, which initially break its physical consistency. Sustaining this measurement frequency prevents decorrelation and progressively suppresses the residual errors, until the temperature is qualitatively assimilated, and the vorticity nearly so, after 50 measurements (\hyperref[fig:measuring_whole_fields]{column (d)}), which, at this observation frequency, correspond to about one large-scale turnover time $t_L$.

Physically, data assimilation proceeds scale by scale: the filter first constrains the large energetic scales, establishing global coherence, after which the small-scale dissipative interactions are continuously tracked and corrected to prevent the system from decorrelating again and drifting away from the ground truth (as decorrelation originates at the faster, smaller scales). This mechanism is visible in \hyperref[fig:corrections_whole_fields]{Fig.~\ref{fig:corrections_whole_fields}(a1) and (a2)}, which shows the relative spectral correction (\hyperref[eqn:rel_corection]{Eq.~(\ref{eqn:rel_corection})}) at all wavenumbers as a function of time in the interval $0<t/\Delta t_{\text{obs}}<400$, after which the performance of the filter remains constant.

Initially, the EnKF corrects the temperature field $\theta^{(n)}$ at all scales before concentrating on the small scales, with occasional corrections rippling back to the large scales. In contrast, the corrections of the vorticity field $\omega^{(n)}$ are predominantly concentrated at small scales, and the large-scale corrections eventually vanish over time, as illustrated in \hyperref[fig:corrections_whole_fields]{Fig.~\ref{fig:corrections_whole_fields}(a2)}. Consequently, both fields are influenced by the data assimilation process and lock onto the ground truth, with the vorticity exhibiting a slight delay as information propagates through the degrees of freedom of the system. The same behavior emerges from the normalized error spectra $E_{\delta\theta}^{(n)}(k,t)$ and $E_{\delta\omega}^{(n)}(k,t)$ (\hyperref[fig:corrections_whole_fields]{Fig.~\ref{fig:corrections_whole_fields}(b1) and (b2)}), which decrease first at the large scales and then progressively at the smaller ones. The error on the measured temperature drops rapidly, and its residual small-scale noise quickly disappears, whereas the error on the vorticity persists longer. The evolution of the errors is thus consistent with that of the corrections (\hyperref[fig:corrections_whole_fields]{Fig.~\ref{fig:corrections_whole_fields}(a1) and (a2)}) and confirms a rapid large-scale reconstruction followed by a progressive multiscale error reduction in both fields.
\begin{figure*}[b]
    \centering
    \includegraphics[width=\linewidth]{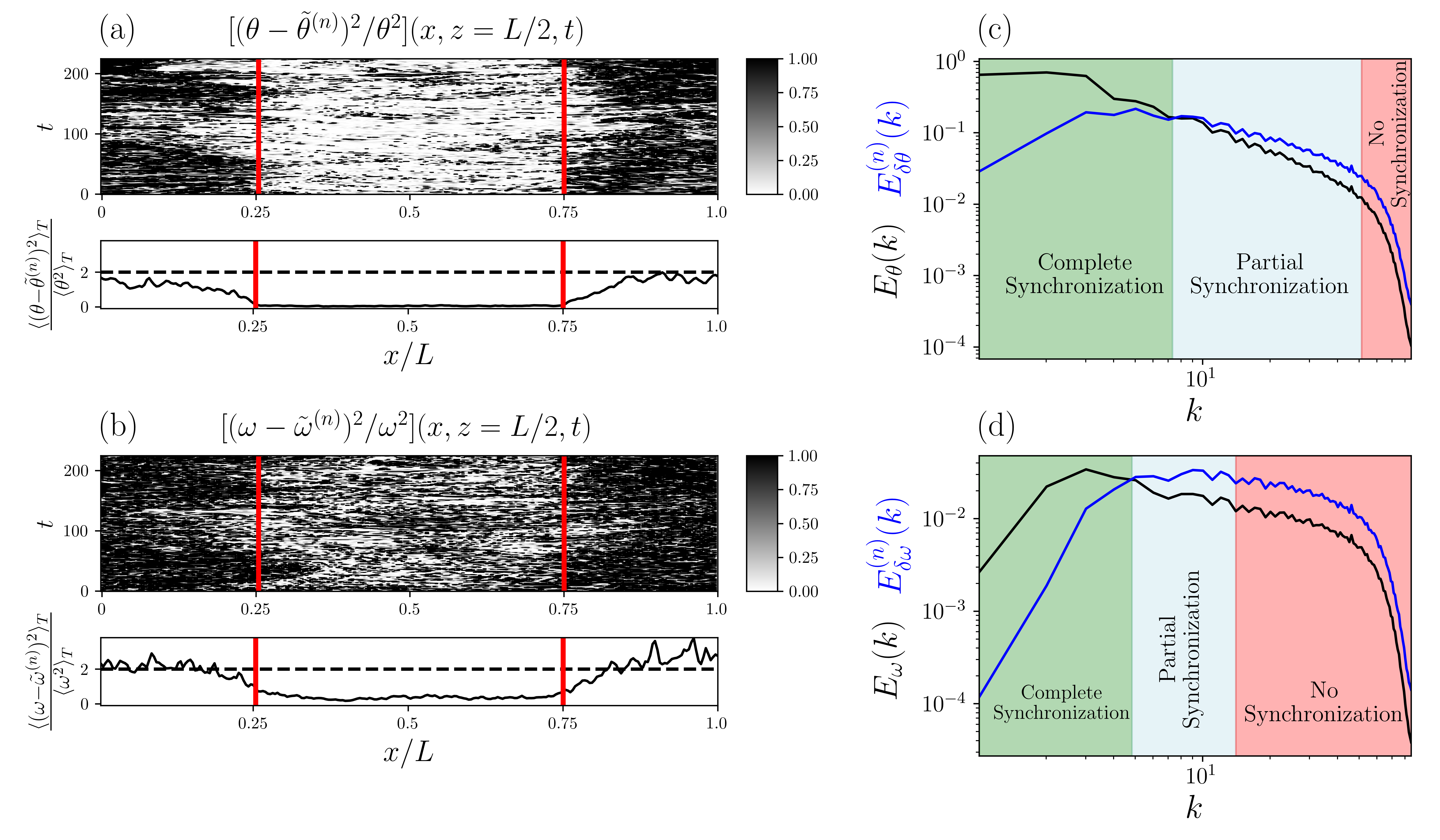}
    \caption{Normalized spatial errors of the temperature (a) and of the vorticity (b) along the line $z=L/2$, when only the central subdomain $0.25<x/L<0.75$, $0.25<z/L<0.75$ of the $256^2$ grid is measured. The upper parts show the instantaneous normalized errors $(\theta-\tilde{\theta}^{(n)})^2/\theta^2$ and $(\omega-\tilde{\omega}^{(n)})^2/\omega^2$ as a function of time, while the lower parts show the ratio of their time averages, $\langle(\theta-\tilde{\theta}^{(n)})^2\rangle_T/\langle\theta^2\rangle_T$ and $\langle(\omega-\tilde{\omega}^{(n)})^2\rangle_T/\langle\omega^2\rangle_T$. The dashed horizontal lines at the value $2$ denote statistical independence, i.e., the spatial limit of the assimilation, while the red lines mark the boundaries of the measured subdomain. Panels (c) and (d) display the time-averaged error spectra in the saturation regime, highlighting three distinct ranges: complete synchronization, partial synchronization, and absence of synchronization.}
    \label{fig:corrections_box_fields}
\end{figure*}
A quantitative assessment over the interval $200<t/\Delta t_{\text{obs}}<400$, where the error has become stationary, confirms the efficacy of the assimilation. This is evident when averaging the spectral errors $E_{\delta\theta}^{(n)}(k)=\langle E_{\delta\theta}^{(n)}(k,t)\rangle_T$ and $E_{\delta\omega}^{(n)}(k)=\langle E_{\delta\omega}^{(n)}(k,t)\rangle_T$ within this time window and comparing them with the mean ground-truth spectra $E_{\theta}(k)=\langle E_{\theta}(k,t)\rangle_T$ (\hyperref[fig:corrections_whole_fields]{Fig.~\ref{fig:corrections_whole_fields}(c1)}) and $E_{\omega}(k)=\langle E_{\omega}(k,t)\rangle_T$ (\hyperref[fig:corrections_whole_fields]{Fig.~\ref{fig:corrections_whole_fields}(c2)}). Indeed, \hyperref[fig:corrections_whole_fields]{panels (c1) and (c2)} show that the error remains below the signal at every resolved wavenumber, for both the measured and the unmeasured field, i.e., $E_{\delta\theta} < E_\theta$ and $E_{\delta\omega} < E_\omega$ for all $k$. Moreover, \hyperref[fig:corrections_whole_fields]{panel (c1)} shows that the error of the EnKF estimate is smaller than the measurement noise.

\subsection{Measuring Temperature in a Subdomain}
\label{sub:box}
To test the framework under more restrictive conditions, a second set of experiments investigates the data assimilation performance when a dense observational network is concentrated exclusively within a subdomain $[0.25\leq x/L\leq 0.75]\times[0.25\leq z/L\leq 0.75]$ located at the center of the domain. Inside the subdomain, all resolved scales are measured, but no direct information is available on the rest of the domain, which covers three quarters of the total area. The observation interval is kept at $\Delta t_{\text{obs}} = t_{\lambda}/8$, and the inflation prefactor, tuned through extensive testing, is set to $\eta = 0.3$, smaller than in the full-field case. This is because the same inflation coefficient is applied to the whole field, whereas the analysis contracts the ensemble more strongly inside the subdomain than outside it, so that an inflation sized for the observed region over-disperses the ensemble in the unobserved one. Indeed, we find that increasing $\eta$ degrades the estimate first outside the subdomain and then also inside it.

\hyperref[fig:corrections_box_fields]{Figure~\ref{fig:corrections_box_fields}} displays the time evolution of the normalized temperature error, $(\theta - \tilde{\theta}^{(n)})^2 / \theta^2$, along the horizontal line $z = L/2$ for a randomly chosen ensemble member (panel (a)), together with the corresponding vorticity error, $(\omega - \tilde{\omega}^{(n)})^2 / \omega^2$ (panel (b)). Within the observed subdomain, the local temperature error drops sharply, driving the local reconstruction of the unmeasured vorticity field through cross-variable coupling. Coherent structures synchronized inside the subdomain are subsequently advected and diffused outward by the flow, temporarily carrying useful information into unobserved regions. However, in the absence of direct measurements outside the subdomain (e.g., for $x/L<0.25$ or $x/L>0.75$ in \hyperref[fig:corrections_box_fields]{Fig.~\ref{fig:corrections_box_fields}(a)}), these structures are progressively decorrelated and destroyed by the chaotic dynamics.

The time-averaged error profiles, shown at the bottom of \hyperref[fig:corrections_box_fields]{panels (a) and (b)}, reveal a spatial transport mechanism: synchronization extends moderately beyond the boundaries of the measured subdomain, eventually degrading to a purely statistical reconstruction near the periodic borders, where the normalized error approaches the saturation value $2$. Desynchronization occurs closer to the subdomain edges for the unmeasured vorticity (\hyperref[fig:corrections_box_fields]{Fig.~\ref{fig:corrections_box_fields}(b)}) than for the temperature.

This spatial breakdown and the destruction of synchronized structures are ultimately governed by small-scale fluctuations. Inspection of the error spectra in \hyperref[fig:corrections_box_fields]{Fig.~\ref{fig:corrections_box_fields}(c) and (d)} shows that complete synchronization, defined by $E_{\delta\theta} < E_\theta$ and $E_{\delta\omega} < E_\omega$, is achieved only at the largest scales. This range is followed by one of partial synchronization, where $E_{\delta\theta}/E_\theta < 2$ and $E_{\delta\omega}/E_\omega < 2$, and, at smaller scales, by a range where the signals are not synchronized. Physically, while large scales possess the coherence length and persistence required to survive advective transport outside the observed subdomain, small scales are dominated by rapid local turnover and chaotic mixing, which destroy information faster than advection can supply it. Thus, localized observations establish a scale-dependent horizon of predictability, where successful state estimation is bounded by the competition between spatial transport rates and chaotic decorrelation timescales.

For the unmeasured variable $\omega$, observations in the subdomain still constrain the dynamics and yield complete synchronization at the largest scales, as shown in panel (d). However, both its complete and partial synchronization ranges are narrower than those of the measured variable $\theta$, as evidenced by the higher relative error $E^{(n)}_{\delta\omega}/E_{\omega}$. This is because, at each analysis, the vorticity, which is never observed, is corrected only indirectly, in proportion to its covariance with the observed temperature, $\hat{\pmb{\Sigma}}_{\omega\theta}$, which enters the Kalman gain (\hyperref[eqn:kalman_gain]{Eq.~(\ref{eqn:kalman_gain})}). At the largest scales, where the two fields remain well correlated, this indirect correction is effective. At smaller scales and away from the observed subdomain, the correlation between temperature and vorticity is expected to be weaker, and its estimate from a finite ensemble noisier, so that the vorticity is corrected less. Both effects restrict the range of scales over which the unobserved field can be recovered. Nevertheless, during the subsequent forecast steps, the dynamics further propagates the information carried by the observations to the whole domain and to the unobserved variable.

\subsection{Measurement Frequency Analysis}
\label{sub:frequency}
\begin{figure*}[t]
    \centering
    \includegraphics[width=\linewidth]{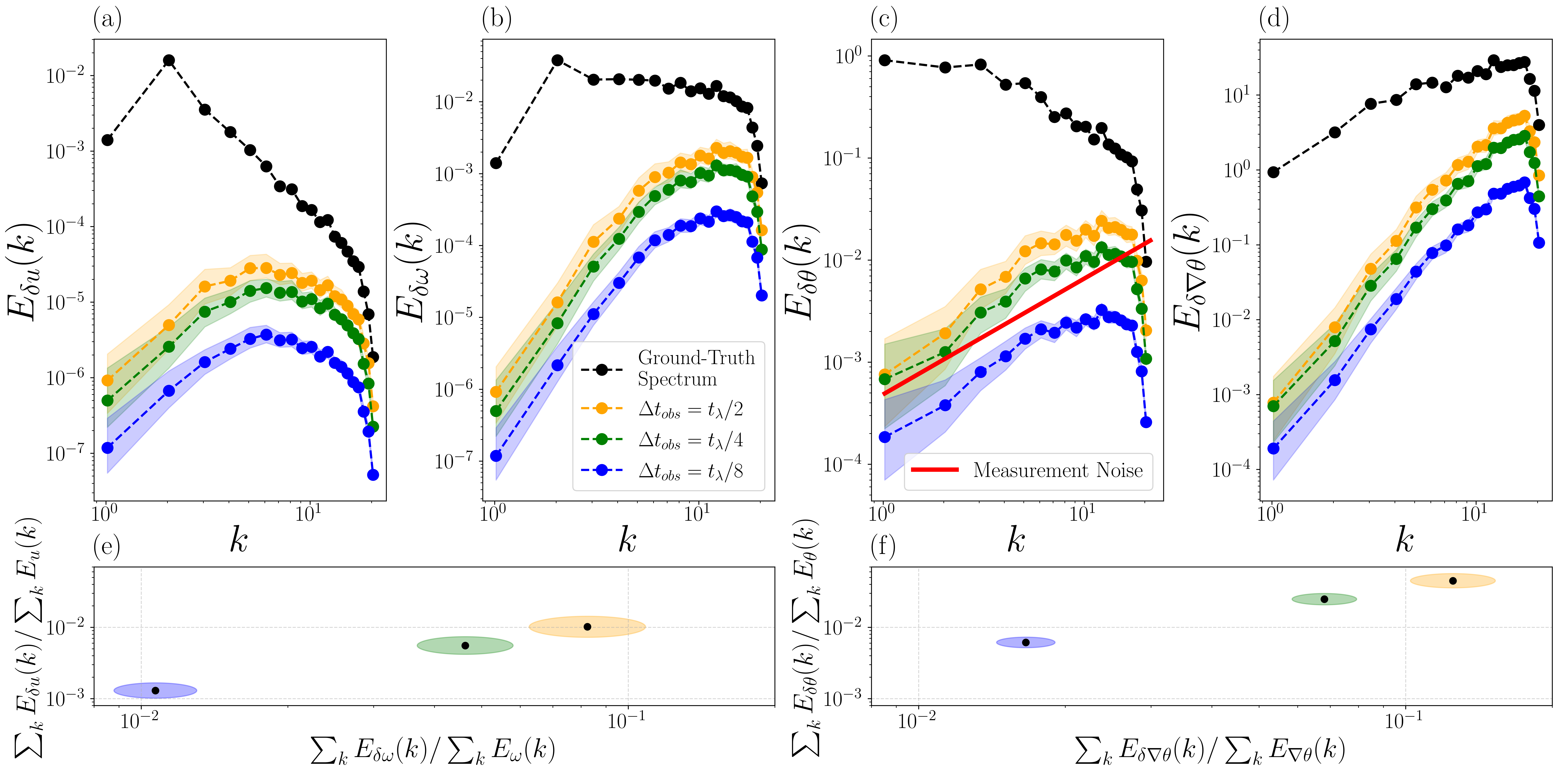}
    \caption{Error spectra on the $64^2$ grid for different observation intervals $\Delta t_{\text{obs}}$, with temperature observations over the whole domain. Shaded regions represent the range between the minimum and the maximum over all the $N_{\text{ens}}$ ensemble members, from $\min_{n} E^{(n)}_{\delta\theta}(k)$ to $\max_{n} E^{(n)}_{\delta\theta}(k)$, while solid lines denote the ensemble mean, $\sum_n E^{(n)}_{\delta\theta}(k)/N_{\text{ens}}$. Results are shown for (a) velocity, (b) vorticity, (c) temperature fluctuations, and (d) temperature gradients. The ground-truth spectra $E(k)$ (black lines) are shown as references for a qualitative assessment of the error-to-signal ratio, e.g., $E_{\delta\theta}(k)/E_\theta(k)$. In panel (c), the red line shows the spectrum of the measurement noise. The spectral panels are complemented by scatter plots comparing the normalized total error of each primary field with that of its spatial derivative: panel (e) compares velocity, $\sum_k E_{\delta u}(k)/\sum_k E_{u}(k)$, with vorticity, $\sum_k E_{\delta\omega}(k)/\sum_k E_{\omega}(k)$, while panel (f) compares temperature, $\sum_k E_{\delta\theta}(k)/\sum_k E_{\theta}(k)$, with its gradient, $\sum_k E_{\delta\nabla\theta}(k)/\sum_k E_{\nabla\theta}(k)$. In the scatter plots, filled markers indicate the ensemble mean, and the semi-axes of the surrounding ellipses represent the ensemble spread, estimated as half the difference between the maximum and minimum total errors over the members, along both axes.}
    \label{fig:frequency_error_spectra}
\end{figure*}

We evaluate the performance of the EnKF for various observation intervals (expressed as fractions of the Lyapunov time) by assimilating full-field temperature observations. To carry out this systematic parameter sweep, experiments are conducted on the $64^2$ configuration, where the lower resolution and, consequently, the smaller number of active degrees of freedom require fewer ensemble members (here $N_{\text{ens}}=2000$), thereby speeding up the numerical simulations and enabling systematic testing. Here, the noise level of the diagonal measurement error covariance is set to $\sigma=1$, again so as to perturb only the high-wavenumber end of the spectrum (see the red curve in \hyperref[fig:frequency_error_spectra]{Fig.~\ref{fig:frequency_error_spectra}(c)}). 
The Lyapunov time is $t_\lambda = 1/\lambda=7.7$. Both the reconstruction accuracy of the measured scalar field and the inferential recovery of the unobserved hydrodynamic state are expected to degrade as the observation interval increases.

In \hyperref[fig:frequency_error_spectra]{Fig.~\ref{fig:frequency_error_spectra}}, we present the error spectra for the scalar field $\theta$, the velocity field $\pmb{u}$, and their spatial gradients, $\nabla\theta$ and the vorticity $\omega$. To establish a scale-dependent signal-to-noise benchmark, these error spectra are compared with the energy spectra of the ground-truth fields. Solid lines depict the ensemble mean over all $N_{\text{ens}}$ members, $\sum_n E^{(n)}_{\delta\theta}(k)/N_{\text{ens}}$, while shaded regions encompass the spread, from $\min_{n \le N_{\text{ens}}} E^{(n)}_{\delta\theta}(k)$ to $\max_{n \le N_{\text{ens}}} E^{(n)}_{\delta\theta}(k)$, quantifying the uncertainty of the reconstruction.
These results show that the EnKF infers the unobserved variables from the observed temperature. For all observation frequencies evaluated in this full-field setup, the filter achieves complete trajectory synchronization ($E_{\delta\theta}(k) < E_\theta(k)$ for all $k$). This also holds for derivative quantities such as $\nabla\theta$ and $\omega = \nabla \times \pmb{u}$. Because small-scale gradients are dominated by intermittency and non-Gaussian heavy tails (reflecting sharp thermal fronts and intense vortical structures), the suppression of errors at these scales shows that the EnKF captures intense, localized events of the dynamics, despite relying on a linear update.

Although all evaluated observation frequencies yield accurate scale-by-scale reconstructions, the EnKF filters out the observational noise (outperforming the raw observations, shown in red) only when the observation interval satisfies $\Delta t_{\text{obs}} \lesssim  t_\lambda / 8$ (see \hyperref[fig:frequency_error_spectra]{Fig.~\ref{fig:frequency_error_spectra}(c)}). In this high-frequency regime, the assimilation cycle intercepts error growth before chaotic small-scale modes contaminate the inertial cascade. Consequently, the reconstruction of the unmeasured fields and of the temperature gradient improves, as shown in \hyperref[fig:frequency_error_spectra]{Figs.~\ref{fig:frequency_error_spectra}(a), (b), and (d)}.

Because spatial gradients probe fine-scale fluctuations, their spectra peak at the small dissipative scales, whereas the primary fields concentrate their energy at large scales. Thus, the normalized total error $\sum_k E_{\delta\theta}/\sum_k E_{\theta}$ is dominated by the large scales, while $\sum_k E_{\delta\nabla\theta}/\sum_k E_{\nabla\theta}$ reflects the small-scale dynamics. Comparing these metrics clarifies where the correction succeeds. As shown in \hyperref[fig:frequency_error_spectra]{panels~(e) and (f)}, where filled markers denote the ensemble mean and the ellipse semi-axes the ensemble half-dispersion of these total errors, i.e., half the difference between their maximum and minimum over the members, the small scales systematically lose controllability as the observation frequency decreases. Unmeasured fields such as the velocity achieve lower relative errors than the temperature, whose recovery remains bounded by the measurement noise. Finally, while the primary fields show a small asymptotic ensemble spread, their gradients show a larger one, reflecting the sensitivity of the small scales to individual realizations.
\begin{table}[b]
\begin{ruledtabular}
    \begin{tabular}{ccccc}
     $\Delta t_{\text{obs}}/t_{\lambda}$ & $N_{\text{meas}}/N^2$ & $\eta$ & $\mathcal{E}_\theta^{\text{tot}}\pm\Delta\mathcal{E}_\theta^{\text{tot}}$ & $\mathcal{E}_u^{\text{tot}}\pm\Delta\mathcal{E}_u^{\text{tot}}$ \\
    \hline
 1/8  & 100\%  & 0.4 & $0.28\pm0.08$ & $0.29\pm0.05$ \\
 1/4 & 100\% & 0.6 & $1.14\pm0.18$ & $1.27\pm0.26$ \\
 1/2  & 100\% & 0.6 & $2.09\pm0.43$ & $2.27\pm0.58$\\
 1/8 & 25\% (equispaced) & 0.4 & $1.24\pm0.19$ & $1.33\pm0.29$ \\
 1/8 & 6.25\% (equispaced) & 0.3 & $5.75\pm0.17$ & $6.12\pm1.60$ \\
 1/8 & 1.56\% (equispaced) & 0.1 & $28.16\pm4.94$ & $29.52\pm5.66$
    \end{tabular}
    \end{ruledtabular}
    \caption{Summary of the analyzed cases on the $64^2$ grid, including the inflation prefactors $\eta$. The total errors are obtained by summing over all resolved wavenumbers the scale-by-scale errors normalized by the ground-truth energy spectra, e.g., $\mathcal{E}_\theta=\sum_k E_{\delta\theta}(k)/E_\theta(k)$ and $\mathcal{E}_u=\sum_k E_{\delta u}(k)/E_u(k)$, and can be obtained from \hyperref[fig:frequency_error_spectra]{Figs.~\ref{fig:frequency_error_spectra}(a)--(d)} and \hyperref[fig:coarsening_error_spectra]{Figs.~\ref{fig:coarsening_error_spectra}(a)--(d)}. Since the spectrum of the gradient of a field satisfies $E_{\nabla\phi}(k)\simeq k^2E_\phi(k)$, the normalized error spectra of the temperature gradient and of the vorticity coincide with those of the temperature and of the velocity, respectively, and are therefore not reported. Tabulated values are ensemble averages, while the uncertainties are the ensemble half-dispersions, defined for the temperature as $\Delta\mathcal{E}_\theta = \big(\max_{n \le N_{\text{ens}}} \sum_k E_{\delta\theta}^{(n)}(k)/E_\theta(k) - \min_{n \le N_{\text{ens}}} \sum_k E_{\delta\theta}^{(n)}(k)/E_\theta(k)\big)/2$, and analogously for the velocity. Since each normalized term equals 2 for an estimate that is statistically consistent with the ground truth but uncorrelated with it, i.e., $E_{\delta}(k) \approx 2E(k)$, a total error of $2\times20=40$ corresponds to the absence of synchronization over all the $20$ resolved scales.}
    \label{tab:measurement_sparseness}
\end{table}
To compare the different configurations with a single number, we define the total normalized errors for the temperature and velocity fields as $\mathcal{E}_\theta^{\text{tot}}=\sum_k E_{\delta\theta}(k)/ E_\theta(k)$ and $\mathcal{E}_u^{\text{tot}}=\sum_k E_{\delta u}(k)/E_u(k)$, averaged over the ensemble, with the uncertainty estimated through the ensemble half-dispersion. With this definition, the contribution of each wavenumber saturates at $2$ when the estimate is uncorrelated with the ground truth ($E_{\delta}(k) \approx 2E(k)$). Consequently, a total error of $2 \times 20 = 40$, i.e., twice the number of resolved wavenumbers, corresponds to a complete loss of skill. The inflation prefactor $\eta$ (see \hyperref[eqn:inflation_coeff]{Eq.~(\ref{eqn:inflation_coeff})}), tuned for each configuration, also depends on the observation frequency, increasing from $\eta=0.4$ for $\Delta t_{\text{obs}}=t_\lambda/8$ to $\eta=0.6$ for the two longer intervals. By construction, the inflation coefficients $g_\theta$ and $g_\omega$ already grow with the spread reduction produced at each analysis, which is larger for longer intervals, since the members drift further from the ground truth, and from each other, between two consecutive observations. This automatic adjustment alone is nonetheless insufficient for the longer intervals, which require a larger prefactor. Note also that $\eta$ controls the inflation applied at each analysis, so that the smaller prefactor used for more frequent observations, being applied at more analysis steps per unit time, does not necessarily result in a weaker cumulative inflation.

Performance is summarized in \hyperref[tab:measurement_sparseness]{Table~\ref{tab:measurement_sparseness}} using the total normalized errors for the temperature and velocity fields, $\mathcal{E}_\theta^{\text{tot}}$ and $\mathcal{E}_u^{\text{tot}}$. The table also reports the inflation prefactor $\eta$ applied in each configuration.

\begin{figure*}[b]
    \centering
    \includegraphics[width=\linewidth]{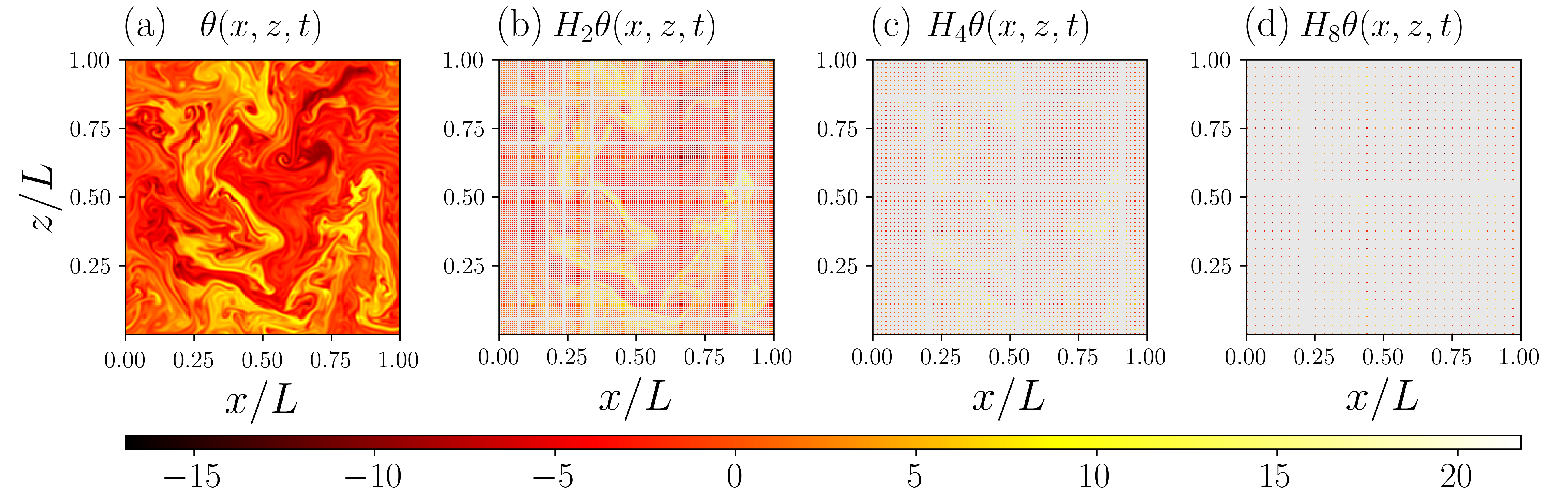}
    \caption{Temperature fluctuation field $\theta(x,z,t)$ at (a) full resolution and after equispaced spatial subsampling by the operator $H_c$, retaining one grid point every $c$ in each direction: (b) $H_2\theta$, (c) $H_4\theta$, and (d) $H_8\theta$.}
    \label{fig:coarse}
\end{figure*}
\subsection{Spatial Sparseness of Measurements}
\label{sub:coarsening}
Having established the impact of the observation frequency, we investigate how the inference performance depends on the spatial resolution of the measurement network, as illustrated in \hyperref[fig:coarse]{Fig.~\ref{fig:coarse}}. To isolate spatial sampling effects, we fix the observation interval at the shortest value considered in the previous analysis, $\Delta t_{\text{obs}} = t_\lambda / 8$, and vary the number of measured temperature points across the domain. Specifically, the temperature field is subsampled on a grid with spacing $c = 1, 2, 4, 8$ grid points, progressively reducing the density of measurements from $100\%$ to $\approx 1.56\%$ of the grid points.

As the coarsening factor $c$ varies, the noise variance $\sigma^2$ of each measured point, i.e., the diagonal entries of $\pmb{R}$, is kept equal to that of the full-grid experiment, while the dimension of $\pmb{R}$ decreases with the number of measurement locations. In principle, spatial coarsening introduces aliasing effects that contaminate the resolved scales with representativeness error~\cite{janjic_2018,tandeo_2020}. Although accounting for this subgrid contribution strictly requires re-estimating $\pmb{R}$, we do not model the representativeness error explicitly and rely on covariance inflation to compensate for it implicitly.
\begin{figure*}[b]
    \centering
    \includegraphics[width=\linewidth]{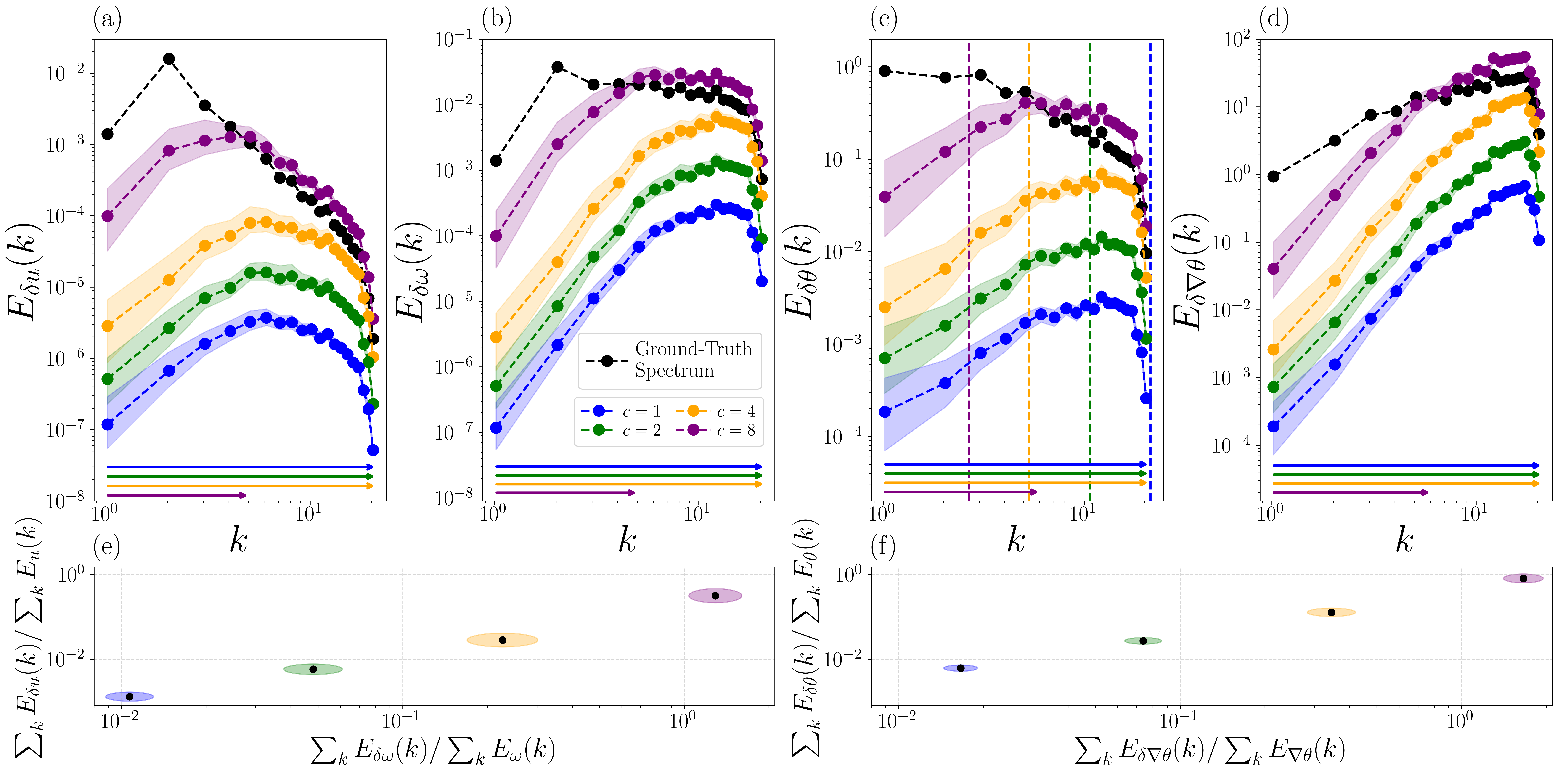}
    \caption{Error spectra on the $64^2$ grid at a fixed observation interval $\Delta t_{\text{obs}}=t_\lambda/8$, for different spatial coarsening factors of the observations ($c=1,2,4,8$). Shaded regions represent the range between the minimum and the maximum over all the $N_{\text{ens}}$ ensemble members, from $\min_{n} E_{\delta\theta}^{(n)}(k)$ to $\max_{n} E_{\delta\theta}^{(n)}(k)$, while solid lines denote the ensemble mean, $\sum_n E_{\delta\theta}^{(n)}(k)/N_{\text{ens}}$. Results are shown for (a) the velocity error $E_{\delta u}(k)$, (b) the vorticity error $E_{\delta\omega}(k)$, (c) the temperature error $E_{\delta\theta}(k)$, and (d) the temperature-gradient error $E_{\delta\nabla\theta}(k)$. The ground-truth spectra $E(k)$ (black lines) are shown as references for a qualitative assessment of the error-to-signal ratio. In panel (c), the dashed vertical lines mark the measured wavenumber range for each coarsening level, which extends from $k=0$ up to the wavenumber of the smallest measured scale. In all panels, the colored arrows indicate the wavenumber up to which complete synchronization is achieved, $E_{\delta}(k)<E(k)$. The spectral panels are complemented by scatter plots comparing the normalized total error of each primary field with that of its spatial derivative: panel (e) compares velocity, $\sum_k E_{\delta u}(k)/\sum_k E_{u}(k)$, with vorticity, $\sum_k E_{\delta\omega}(k)/\sum_k E_{\omega}(k)$, while panel (f) compares temperature, $\sum_k E_{\delta\theta}(k)/\sum_k E_{\theta}(k)$, with its gradient, $\sum_k E_{\delta\nabla\theta}(k)/\sum_k E_{\nabla\theta}(k)$. In the scatter plots, filled markers indicate the ensemble mean, and the semi-axes of the surrounding ellipses represent the ensemble spread, estimated as half the difference between the maximum and minimum total errors over the members, along both axes.}
    \label{fig:coarsening_error_spectra}
\end{figure*}
All other system parameters and the initialization protocol are unchanged with respect to the previous experiments. Performance is evaluated through the error spectra $E_{\delta\theta}^{(n)}(k)$, compared with the ground-truth spectra $E_{\theta}(k)$. This setup isolates the sensitivity to the spatial information content and reveals the resolution threshold below which the EnKF can no longer reconstruct the full spectral range of the fields.

The top panels of \hyperref[fig:coarsening_error_spectra]{Fig.~\ref{fig:coarsening_error_spectra}} show that complete scale-by-scale synchronization ($E_{\delta\theta}(k) < E_\theta(k)$) is maintained for both the measured and the unmeasured field, as well as for their spatial gradients, up to $c = 4$, which corresponds to measuring $6.25\%$ of the grid points. Even with partial spatial observations, the cross-covariances thus propagate information across scales, and the small-scale, non-Gaussian gradient fields $\nabla\theta$ and $\omega$ are correctly inferred.

Unlike in the observation-frequency experiments, where synchronization over the full spectrum was achieved in all cases, the experiments in spatial coarsening reveal a threshold at $c = 8$. At this coarsening level (fewer than $2\%$ of the grid points measured), accurate reconstruction is confined to a narrow low-wavenumber band (indicated by the purple arrow). Beyond this range, the spectral error grows rapidly and saturates at $E_{\delta\theta}(k) \approx 2E_\theta(k)$. This factor of $2$ signals that, at high wavenumbers for $c = 8$, the reconstructed state and the ground truth are statistically independent. At these unobserved small scales, the EnKF loses phase coherence with individual turbulent eddies, assimilation fails, and the state reconstruction degrades to a purely statistical match.

\hyperref[fig:coarsening_error_spectra]{Figures~\ref{fig:coarsening_error_spectra}(e) and \ref{fig:coarsening_error_spectra}(f)}, where filled markers denote the ensemble mean and the ellipse semi-axes the ensemble half-dispersion of the total errors, i.e., half the difference between their maximum and minimum over the members, show the relation between the normalized total errors of the primary fields and those of their spatial gradients. As in the observation-frequency analysis, this relation follows a linear trend, with the performance degradation governed by the coarsening level and by the sensitivity of the gradients. For $c = 8$, the normalized total errors are of order unity, confirming that synchronization remains confined to the largest energetic scales.

As reported in \hyperref[tab:measurement_sparseness]{Table~\ref{tab:measurement_sparseness}}, the inflation prefactor is unchanged between $c=1$ and $c=2$ ($\eta=0.4$) and decreases for stronger coarsening, to $\eta=0.3$ for $c=4$ and $\eta=0.1$ for $c=8$. We attribute this trend to the same mechanism discussed for the subdomain case (see \hyperref[sub:box]{Subsec.~\ref{sub:box}}), now acting at the scale of the observation spacing. With an equispaced network, the analysis reduces the ensemble spread more strongly at the measured grid points and progressively less at the points in between, which are corrected only through their covariances with the observed ones. Since the inflation coefficients are scalars applied uniformly to all grid points, an inflation sufficient to compensate the spread reduction at the measured points over-disperses the ensemble at the unmeasured ones. As $c$ increases, the fraction of unmeasured points grows and their distance from the nearest observation increases, which makes the mismatch larger and, in turn, requires a smaller $\eta$. In spectral space, the unmeasured points correspond to the small scales not resolved by the network, consistently with the restriction of synchronization to the largest scales observed at strong coarsening.

\section{Conclusion}
\label{sec:conclusion}
In this paper, we have shown that the Ensemble Kalman Filter can reconstruct multiscale turbulent convection from temperature observations alone, thereby demonstrating that there exist an observational scenario and a data assimilation framework in which Charney's conjecture~\cite{charney_use_1969} holds. A key limitation of DR and nudging is the requirement to identify and directly measure the determining modes of the system to achieve complete trajectory synchronization~\cite{carlson2024,patricio,pazo_data_assimilation_2016}. This introduces an operational asymmetry: measuring non-determining modes (such as the temperature in this case, see \hyperref[app:direct_replacement]{Appendix~\ref*{app:direct_replacement}}, while in general it is not clear a priori how to isolate the determining modes) fails to trigger synchronization and thus assimilation~\cite{titi_victor2025}. By contrast, the EnKF removes this asymmetry by exploiting flow-dependent cross-correlations, bypassing the need for explicit mode selection and directly inferring the unobserved fields across the entire multiscale spectrum. In doing so, the EnKF overcomes the limitations of conventional synchronization-based techniques, such as DR and continuous nudging, and offers an alternative to Physics-Informed Neural Networks (PINNs)~\cite{di_leoni_pinns_rayleigh_2023}, which can suffer from spectral bias or phase misalignment when attempting to recover non-Gaussian gradient fields without dense spatio-temporal supervision.

These results are demonstrated in two distinct chaotic regimes, both representative of fully developed multiscale turbulence with intermittency and an extended inertial range. In the more chaotic $256^2$ system, measuring the entire temperature field enables the filter to achieve full scale-by-scale trajectory synchronization. The EnKF reconstructs the complete velocity field $\pmb{u}$ and high-pass quantities such as the vorticity $\omega$, capturing localized extreme events. Measuring the temperature at all grid points amounts to measuring all its resolved scales. Whereas a dense network of measurements provides complete spectral knowledge, measurements concentrated inside a subdomain allow accurate tracking of the ground truth only within the observed region; the synchronized coherent structures are then transported by the flow, inducing a synchronization wave that dissipates outside the subdomain within a decorrelation time, so that only the large scales remain synchronized. In the less chaotic $64^2$ regime, systematic parameter sweeps reveal the operational bounds of this multiscale synchronization: for sufficiently frequent observations ($\Delta t_{\text{obs}} \lesssim t_\lambda / 8$), the estimate becomes more accurate than the measurements themselves, while a spatial density of measurements as small as $6.25\%$ was found to be sufficient to preserve phase coherence across the unobserved small scales.

Although these findings establish the capability of the EnKF to exploit cross-covariances across a broad range of spatial and temporal scales, its operational applicability remains restricted to relatively low-dimensional models because of the curse of dimensionality. In realistic, high-Reynolds-number applications, such as operational weather forecasting, resolving the large unstable subspace with explicit ensembles is computationally prohibitive~\cite{carrassi_chaotic_dynamics_2022}. These limitations motivate the exploration of alternatives, including machine learning approaches, to complement~\cite{cheng_ml_da_uq_2023}, partially replace, or fully supersede the EnKF and more general data assimilation frameworks. Promising directions include learning covariance corrections for small-ensemble regimes by leveraging the dynamical history~\cite{solver_in_the_loop,freitas2025} or optimizing the initial ensemble configuration to maximize variability~\cite{shin_diffusion_2026}.


\begin{acknowledgments}
The authors acknowledge useful discussions with Fabio Bonaccorso (University of Rome ``Tor Vergata'').

This research is supported by the European Union's HORIZON MSCA Doctoral Networks programme under Grant Agreement No.~101072344, project AQTIVATE (Advanced computing, QuanTum algorIthms and
data-driVen Approaches for science, Technology and Engineering), by the European Research Council (ERC) under the European Union's Horizon 2020 research and innovation programme Smart-TURB (Grant Agreement No.~882340), and by the MUR-FARE project R2045J8XAW. AC has been funded as a member of the Scale-Aware Sea Ice Project (SASIP), supported by grant G-24-66154 of Schmidt Sciences, LLC---a philanthropy that propels scientific knowledge and breakthroughs towards a thriving world.
\end{acknowledgments}

\appendix
\section{Data Assimilation via Direct Replacement}
\label{app:direct_replacement}
The understanding of synchronization in dissipative dynamical systems through direct substitution originates from the seminal work of Pecora and Carroll~\cite{pecora_1990}, who demonstrated how coupling a reference dynamical system to a slave system via specific driving variables can induce stable synchronization. In the context of dissipative fluid systems, this paradigm is deeply connected to the theory of determining modes~\cite{foias_sur_1967,olson_titi_determining_modes}, which suggests that the asymptotic behavior of a chaotic attractor is uniquely determined by a finite number of low-wavenumber modes.

To assess the synchronization efficiency, quantified by the convergence rate of the error dynamics, under constraints imposed by either the temperature field $\theta$ or the vorticity field $\omega$, we linearize the governing~\hyperref[eqn:RB]{equations~(\ref{eqn:RB})} around a reference state $(\theta, \omega)$, which yields the following equations for the perturbations $(\delta \theta, \delta \omega)$:
\begin{equation}
\left\{
\begin{aligned}
\label{eqn:perturbations}
\partial_t(\delta \theta) &+ (\delta \pmb{u} \cdot \nabla) \theta + (\pmb{u} \cdot \nabla) \delta \theta &&= \delta u_z \gamma + \kappa\nabla^{2} \delta \theta \\
\partial_t(\delta \omega) &+ (\delta \pmb{u} \cdot \nabla) \omega + (\pmb{u} \cdot \nabla) \delta \omega &&= g\alpha \partial_x (\delta \theta) 
+ \nu\nabla^{2}\delta \omega - \mu\,\delta \omega &&
\end{aligned}
\right. .
\end{equation}

For direct replacement, two cases can be considered: either only the vorticity is known, i.e., $\delta \omega=0$ and $\delta \pmb{u}=0$ (indeed, in 2D incompressible flows knowing $\omega$ is equivalent to knowing $\pmb{u}$), or only the temperature is known, i.e., $\delta \theta=0$. In the first case, the error dynamics~(\ref{eqn:perturbations}) reduce to an advection-diffusion equation; multiplying by $\delta\theta$ and integrating over the domain $\Omega$, i.e., taking the $L^2(\Omega)$ inner product with $\delta\theta$, we obtain
\begin{equation}
\label{eqn:theta_pert_energy}
\!\frac{d}{dt} \left( \frac{1}{2} \| \delta \theta \|_2^2 \right)\!+\! \int_{\Omega} \delta \theta (\pmb{u} \cdot \nabla) \delta \theta \, d\pmb{r} \!=\! \kappa\! \int_{\Omega} \delta \theta \nabla^{2}\delta \theta \, d\pmb{r}\ .
\end{equation}
Owing to incompressibility and the divergence theorem, the advection term vanishes for periodic boundary conditions or for perturbations that vanish at the boundaries. Integrating the right-hand side of \hyperref[eqn:theta_pert_energy]{Eq.~(\ref{eqn:theta_pert_energy})} by parts, we obtain
\begin{equation}
\frac{d}{dt} \left( \frac{1}{2} \| \delta \theta \|_2^2 \right) = -\kappa \| \nabla\delta \theta \|_2^2 \leq 0\ ,
\end{equation}
which implies that the temperature error asymptotically vanishes. Physically, the error evolves as a decaying passive scalar.

In the second case, where the system is driven by the temperature, we set $\delta \theta = 0$ in \hyperref[eqn:perturbations]{Eq.~(\ref{eqn:perturbations})}; an energy argument similar to the previous one shows that the vorticity error evolves as
\begin{equation}
\begin{aligned}
\frac{1}{2} \frac{d}{dt} \| \delta \omega \|_2^2 + \int_{\Omega} \delta \omega (\delta \pmb{u} \cdot \nabla) \omega \, d\pmb{r} = -\nu \| \nabla \delta \omega \|_2^2 - \mu \| \delta \omega \|_2^2
\end{aligned}\ .
\end{equation}
Here, the integral term represents the interaction between the vorticity perturbation and the background vorticity gradient, mediated by the induced velocity perturbation $\delta \pmb{u}$. Unlike in the previous case, this term is not identically zero, and its magnitude is controlled by $\|\nabla \omega\|_\infty$. While the dissipative terms dominate in non-chaotic regimes, ensuring the convergence of $\delta \omega$ to zero, in chaotic regimes this term has no definite sign and may act as an energy source that precludes synchronization. This theoretical limitation is mirrored by numerical findings on \textit{nudging}, where attempts to infer the vorticity exclusively from the temperature fail in chaotic regimes in the absence of additional constraints~\cite{agasthya_rayleigh_2022}. 

These considerations hold equally in the presence of hyper-viscosity, hyper-diffusion, and hypo-friction of arbitrary order.

\bibliographystyle{apsrev4-2}
\bibliography{main}

\vspace{0.1cm}
\hypertarget{emailnote}{\textsuperscript{\textcolor{blue}{*}}\href{mailto:ffossella3@gmail.com}{ffossella3@gmail.com}}

\end{document}